\documentstyle[preprint,aps]{revtex}  
\input psfig.tex
 
\begin{document}
 \draft
 \title{Topological model of soap froth evolution with deterministic
T2-processes.}
 \author{Boris Levitan$^1$ and Eytan Domany$^{1,2}$}
 \address{$^1$Department of Physics of Complex Systems, Weizmann Institute
of Science, Rehovot, Israel}
\address{$^2$Department of Physics, Hong Kong University of Science and
Technology,
Clearwater Bay, Kowloon, Hong Kong}
\date{\today}
\maketitle
\begin{abstract}
We introduce a modified topological model for the evolution of two-dimensional
soap froth.  The topological rearrangement associated with a
T2 process is deterministic; the final topology depends on the
areas of the neighboring cells. The new model gives agreement with experiments
in the transient regime, where the previous models failed qualitatively, and
also improves agreement in the scaling state.
\end{abstract}
\narrowtext
\vskip2pc
Dynamics of soap froths has attracted
considerable attention among physicists\cite{Reveiw}. Evolution of a
two dimensional froth is governed by the
von Neumann law\cite{VonN}:
\begin{equation}
{da_n\over dt}=k(n-6),
\end{equation}
where $a_n$ is the area of any  $n$-sided bubble.
The growth rate of a bubble depends only on its
topological class (TC - the
number of its sides), and is completely independent of any other detail,
such as the TC of its neighbors.
As the froth evolves, bubbles with $n< 6$ sides shrink
and disappear: when that happens,
sudden {\it topological rearrangements}
(called T2 processes) take place.
In the course of these processes the froth regains co-ordination number 3,
and some bubbles that were neighbors of the one that disappeared may loose or
gain a side. A different topological rearrangement, called T1 process, occurs
by {\it side switching}, without
elimination of any bubble.
The total number of cells in the system decreases with time;
hence the average area of the cells $\bar a$
increases and the froth {\it coarsens}.

A second important finding concerning froth evolution was discovered
experimentally\cite{Smith,Gross,StGl,Stavans}:
after a transient period the system reaches a {\it universal scaling state}.
Scaling, combined with the von Neumann law
imposes that
${\bar a} \sim t$ and that
the topological distribution function is constant in the scaling regime,
$x_n(t)=N_n(t)/N(t)\rightarrow x_n^*$.
The scaling state is universal in that it is independent of the froth's
initial state, the gas used, etc. These results were verified\cite{Stavans},
other properties of the scaling state were measured\cite{Fr85a,Seg}
and
some aspects of the transient period that precedes the scaling
regime were also studied\cite{StGl}.

Both analytic and numerical methods have been used to address froth evolution
theoretically. The analytic (mean-field) studies\cite{Seg,Mar,Fr85b,Flyvb,SDM}
yield, in general, surprisingly good agreement with experimentally
measured distribution functions {\it in the scaling state}. Recently
the same methodology was used to predict some more complicated aspects
of froth dynamics, such as the distribution of {\it survivors}\cite{Surv}
in the scaling state, and the evolution of the topological distribution
function from some special initial states\cite{MyLett,Comment}
For these more delicate and complex properties
agreement with experiment is much weaker and only qualitative.

Surprisingly, we found that rather sophisticated
topological simulation methods\cite{MyLett,Frt,Fr} are as bad or
{\it even worse} in predicting
the same characteristics.
The aim of the present paper is to elucidate
the physical reason for  these discrepancies,
and to present a simple modification of the simulations that
improves agreement with experiment most significantly.

In one of the earliest simulations
Weaire and Kermode\cite{WKerm} described
the froth in terms of the coordinates of the vertices and the gas pressures
in the bubbles. By solving at each time the equations of mechanical equilibrium
for all vertices, they determined the spatial configuration of the
froth. The necessary computation is, however,
so involved that one can not treat a large
number of bubbles (they had only 500 bubbles initially -
subsequently Herdtle and Aref simulated 1024 bubbles on a Cray\cite{Her}).
Kawasaki and co-workers\cite{Nagai} solved
a system of differential equations
for the coordinates of the vertices. They were
able to do handle  up  to 24 000 cells; their
results seem to be in good agreement with experiments.
The von Neumann law holds, however, only approximately for this method.
Potts model simulations\cite{Potz} are also severely
restricted in the number of bubbles that can be treated.

The importance of different aspects of the microscopic dynamics
and their separate effects on the macroscopic evolution of the froth are
most easily studied
by analysing a hierarchy of more and more  simplified {\it topological} models,
identifying which aspect is essential in order to obtain agreement with
increasingly delicate experimental observations.

We turn now to describe and consider carefully the simplest
topological model for soap froth dynamics, developed and used extensively by
many investigators \cite{Surv,MyLett,Frt,Fr}, which we call
"model {\bf A}".

{\bf Model A:}
Rather than following locations of the vertices and cell
boundaries, one keeps track of the cells' {\it areas} $a_i, i=1,2,..N$,
and their
{\it topological connectivity}\cite{Frt,Fr}.
Topology is contained in an $adjacency$ matrix $T_{ij}$, which is set
to be 1 if cells $i$ and $j$ are neighbors, and 0 otherwise\cite{stor}.
Temporal evolution is represented by integrating the von Neumann equations (1)
until an area reaches the value $a_i=0$, corresponding to disappearance of
cell $i$. At this point a local rearrangement of the adjacency matrix takes
place, choosing with equal probability one
out of all the possible side assignments of this T2 process. Thus, even
though the precise geometry of the cells is neglected by this approach,
topological correlations and the von Neumann law for temporal evolution
are taken into account explicitly.

In general our simulations start with
an ordered hexagonal lattice, with slightly
randomized areas (i.e. set
$a_i=1+\zeta_i$ where $\zeta_i <<1$ are taken from some distribution).
A significant number of T1 transformations is performed;
starting from the initial
state obtained this way, and letting the froth evolve according to
model {\bf A}, we have demonstrated the following:
\begin{description}
\item[(a)]
All initial assignments of areas evolve to
the same scaling state.
\item[(b)]
The topological class
distribution was found to be
close to the experimental one, but slightly narrower; whereas the experimental
value of the second moment of this distribution was $\mu_2^{expt}=1.4\pm 0.1$,
simulations gave $\mu_2^{A} = 1.2$.
\item[(c)]
The predicted topological distribution of the {\it survivor} cells
 differed from the measured one.
\end{description}

Regarding result (b), to our surprise mean field theory\cite{Seg} fits the
experiments better, yielding $\mu_2^{MF} \approx 1.4$.
To understand the effect of topological correlations (neglected by mean field),
note that according to  Aboav's law\cite{Aboav}
a cell with a small $n$ is more likely to be the neighbor of one
with a large $n$. Since "topological dynamics",
i.e. change of TC, is caused
mainly by the disappearance of $n<6$-bubbles, clearly cells
with more sides are more topologically mobile than those with small $n$.
As this mobility prevalently reduces $n$, mean field, that
neglects this (correlation-induced) dependence of mobility on $n$,
will overestimate the number of large-sided bubbles in the (topological) steady
 state, thus overestimating $\mu_2$. Hence
the fact that  $\mu_2^{A} < \mu_2^{MF}$ is no surprise, but we did
expect the simulation result to be the one that agrees better with experiment!
Clearly some physically important aspect of the evolution
(mentioned briefly in \cite{Nagai,Fr93b}
is not included  in model {\bf A}.

Important hints regarding this missing aspect came from simulations of
froth evolution that start from {\it nearly ordered initial
conditions}\cite{StGl,Lei}
with only a few, isolated defects perturbing an ideal network of hexagonal
cells.
A defect is produced in the simulations
by performing a single T1 switch, replacing four
adjacent hexagons by two pentagons and two heptagons.
Each defect serves as a nucleation center for the growth of a
cluster of non-hexagonal bubbles. Initially all the clusters evolve
independently; when they become large and meet each other, they start to
interact. At that time the ordered regions vanish and
the froth evolves to the completely disordered, universal scaling state.

The temporal evolution of the second moment $\mu_2(t)$, as measured
for these initial conditions\cite{StGl}, shows that the (initially very narrow)
 distribution broadens and $\mu_2$
increases; finally the scaling state value of 1.4 is obtained. The approach to
this value is, however, non-monotonous!
$\mu_2(t)$ was demonstrated\cite{StGl} to exhibit
a remarkable $peak$ value of $\mu_2\approx 2.6$.
Microscopic\cite{WKerm} simulations\cite{Lei}
of evolution from a similar initial state, working with only
about $500$ cells,
reproduced such a peak, but with the lower value
of $\mu_2\approx 1.9$.

We performed extensive simulations of evolution from similar initial
conditions.  Using model {\bf A}, we never succeeded to produce a
peak; $\mu_2$ grew monotonously (up to fluctuations that were smoothed out by
averaging the data over several runs - see Fig. 1).
A related deficiency of model {\bf A} became evident when results of a recent
study of the {\it single cluster problem}\cite{MyLett} were re-analyzed.
This study showed by a mean field calculation and by simulations (using model
{\bf A}), that
the topological distribution of the bubbles that belong to  the growing cluster
approaches a fixed form when the cluster becomes large. Weaire\cite{Comment}
noticed that the width of this distribution\cite{MyLett}, $\mu_2=0.72$, is too
small to be consistent with the experimentally obtained large peak of $\mu_2$.
Indeed, as anticipated by Weaire, our subsequent simulations of a froth with
isolated defects yielded only a monotonously growing $\mu_2$. This striking
failure of model {\bf A} prompted us to look closely also at the results
(b) - (c) listed above, to identify  the element that is missing from this
model, and to eliminate the discrepancies by incorporating it in our
topological simulations.

Model {\bf A} uses the exact von Neumann equations on a topologically correct
network of cells; hence this phase of the evolution is treated exactly.
An uncontrolled approximation (made when bubbles disappear) is that the
topological rearrangement that takes place occurs {\it at random}, selecting
one of all possible outcomes with {\it equal probability}.

The assumption of
random topological rearrangements is not too bad for the scaling regime,
where the froth is macroscopically isotropic and homogeneous, but breaks down
in the transient period of the evolution from the initial state with isolated
defects. In this period the froth contains fairly isolated
clusters, embedded in an inert region of small hexagons. The interior of these
clusters contains large bubbles with many sides, while their boundaries
contain a relatively high concentration of small $n<6$ cells
\cite{StGl}. Typically, one of the neighbors of such a cell with $n<6$
is a very large bubble, while  the other
$n-1$ are small cells. Photographs of the froth shows that
the edge shared with the large cell is nearly always much longer than the
edges shared with the small cells. As a result of this, the cell with the
large area {\it rarely loses a side} during the topological rearrangement
that follows the disappearance of the $n<6$ cell. This correlation between
the areas of the neighbors and the topological rearrangement that takes place
in a T2 process was completely neglected by model {\bf A}. These
correlations can have a considerable effect when the areas of the neighbors
of a disappearing cell
are significantly different\cite{Reply}. Since area is correlated with the
number of sides\cite{Reveiw,Seg} (Lewis' law)
the effect described above will reduce the rate at which cells with high $n$
loose sides, which, in turn, will increase their number and broaden the
topological distribution. To incorporate these correlation we now
examine closely the manner in which a T2 process occurs.
It has been argued recently\cite{Fr93b} that when a disappearing
bubble becomes much
smaller than its neighbors, its shape is retained while shrinking.
Therefore the identity of the shortest side is preserved
till the moment when its length becomes comparable to the width of a vertex.
At this point
the potential barrier for a T1 switch of the shortest side decreases
significantly and the froth immediately lowers its energy by a T1 process.
Thus we postulate that a T2 process for pentagons starts
with the disappearance of the {\it shortest
edge}  by a T1 process, turning
a pentagon first into a rectangle, which again executes a
T1 switch of its
shortest edge and turns into a triangle, which
simply shrinks to a point. A sequence of these processes is illustrated in
Fig. 2. A similar way of executing T2-processes has been used by
Weaire and Kermode\cite{WKerm}.
This procedure
{\it eliminates all stochasticity} from the froth's evolution: once the initial
state is set, evolution is completely deterministic.

In order to include the modified T2 process described above in a numerical
algorithm, we need information regarding the length of the edges $l_{i,j}$
separating two neighboring cells $i$ and $j$.
To extract this information from our {\it topological} model, we looked for a
way to relate the lengths, $l_{i,j}$ to
those properties of the bubbles that were followed anyway in our simulation,
namely the cells' connectivity and areas.
To estimate $l_{i,j}$
consider a
typical bubble $i$ with $n_i$ sides.
Its perimeter $P_i$ is related to its area $a_i$ as $P_i^2\sim a_i$;
assuming then, as first approximation, that $l_{i,j} \sim P_i/ n_i$, we have
$a_i\sim (n_il_{i,j})^2$ which, in turn, implies $l_{i,j}\sim \sqrt{a_i}/n_i$.
On the other hand, the side $l_{i,j}$ belongs to the cell $j$ as well, so that
$l_{i,j}\sim \sqrt{a_j}/n_j$. Therefore,
$l_{i,j}=x_{i,j}\sqrt{a_i}\sqrt{a_j}/n_in_j$, where $x_{i,j}$
depends on the forms of the cells. Our approximation is now to assume that the
$x_{i,j}$ do not vary too much from cell to cell and from side to side, i.e.
$x_{i,j}\approx x_0$, where $x_0$ is the same for all sides in the froth.
Since we need only estimates of the {\it relative} sizes of the sides of a
cell, we hope that the error of our rough estimate is not critical. The factor
$x_0$ can be dropped, and we can rewrite the result in the following final
form:
\begin{equation}
l_{i,j}={\sqrt{a_i}\sqrt{a_j}\over n_in_j}.\label{length}
\end{equation}
This prescription suffices to construct our modified, deterministic topological
model:

{\bf Model  B} differs from model {\bf A} only in the way the
T2-process is performed. When a shrinking bubble reaches a critical size we
calculate the $lengths$ of its sides, using (\ref{length}), and perform a T1
switch of the minimal side, until the shrinking cell is a triangle, which is
reduced to a point.

$\mu_2(t)$, obtained from simulations that start with an
ordered array of hexagons of slightly randomized
areas,
perturbed by a small
concentration of randomly scattered defects,
is presented in Fig. 1. The main qualitative
distinction between the results of the two models is the existence of the
pronounced peak. In addition, the value of
$\mu_2$ in the scaling regime is now also in
agreement with the experimentally obtained  $\mu_2=1.4\pm 0.1$.
Our success
in eliminating these discrepancies between experiment and the previously used
topological model indicates that we have indeed identified correctly the source
 of the disagreement. On the other hand, it should be noted that the magnitude
of the peak, $\mu=1.76$, differs from the experimental value $\mu_2=2.6$, but
is rather close to 1.9, the result of Weaire
and Lei\cite{Lei}. Apparently these two approaches have a common reason for the
discrepancy, and the difference between our simulations and the experiment
is not due to the approximation implicit in the purely topological nature
of our approach.

At early stages when the froth consists of well-separated well developed
clusters, there are very large differences between the areas of the
cells in the clusters' interior versus perimeter
and exterior.
The control parameter
$\sqrt{a}/n$ is much larger for the large bubbles, and the
modified rearrangement
rules have a considerable effect.
In the scaling state of the froth, when there are no
such huge differences between the areas of the neighbors, the
control
ratio is
nearly constant for all $n$ that occur with significant density\cite{Long}.
Therefore in the scaling state
the various topological
rearrangements
occur more or less randomly with equal probabilities, and
model {\bf A} describes the scaling regime rather well. Nevertheless, the
the control ratio still increases slightly\cite{Long}
with $n$, making bubbles with large $n$ less likely
to loose a side when model {\bf B} is used, which explains why
$\mu_2^B \approx 1.4>1.2 \approx \mu_2^A$.

We used model {\bf B} on two other problems
where model {\bf A} has failed. First, we simulated the
single cluster problem.
We found that as in our previous study (that used model {\bf A}),
the topological distribution function of the growing  cluster
approaches a fixed form. As one could expect, model
{\bf B} gives a broader distribution than model {\bf A};
$\mu_2^A \approx 0.7 < 1.1 \approx
\mu_2^B$.

Lastly, we rerun simulations of a large number of cells, and measured again
the topological distribution of the {\it survivor} population\cite{Surv}.
Preliminary results indicate that the distribution obtained using model {\bf B}
is in excellent agreement with experiment, whereas that of model {\bf A} was
significantly off\cite{Long}.

Note that mean field neglects {\it all} correlations - topological, as well
as those introduced here (that depend on the areas).
As discussed earlier, topological correlations tend to increase the rate at
which cells with large $n$ shed their sides (sharpening the topological
distribution), whereas the area-dependent correlations reduce this rate
(broadening it). Thus the two effects act with opposite signs,
which explains why mean field gave, in the scaling state, a distribution
whose width was closer to experiment than that obtained by model {\bf A}.

In summary, we demonstrated that that the manner in which
the topological rearrangement that follows a T2 process is performed
has a strong effect on the dynamics.
The previously used natural choice of selecting the rearrangement at random,
assigning equal probability to every possible outcome, gives sometimes
qualitatively incorrect results. In real froth dynamics the outcome
of the T2-process is strongly correlated with the areas of the neighbors of the
 disappearing bubble. The effect of these correlations is especially
pronounced in the transient period of the evolution from a highly ordered
initial state. In this case the experiments show a remarkable peak of
$\mu_2(t)$, the width of the topological distribution function, which can not
be reproduced
without taking into account the area-dependence of the T2-processes.
We succeeded to incorporate these correlations in our topological model without
making it more involved computationally. The results of the new model agree
much
better with a variety of experimental results than the previous model.

This research was supported by grants from the Germany-Israel Science
Foundation (GIF) and by the US-Israel Binational Science Foundation  (BSF).
B.L. thanks the Clore Foundation for financial support.
We thank D. Weaire for most helpful correspondence and J. Stavans for
discussions.

\begin{figure}
\centerline{\psfig{file=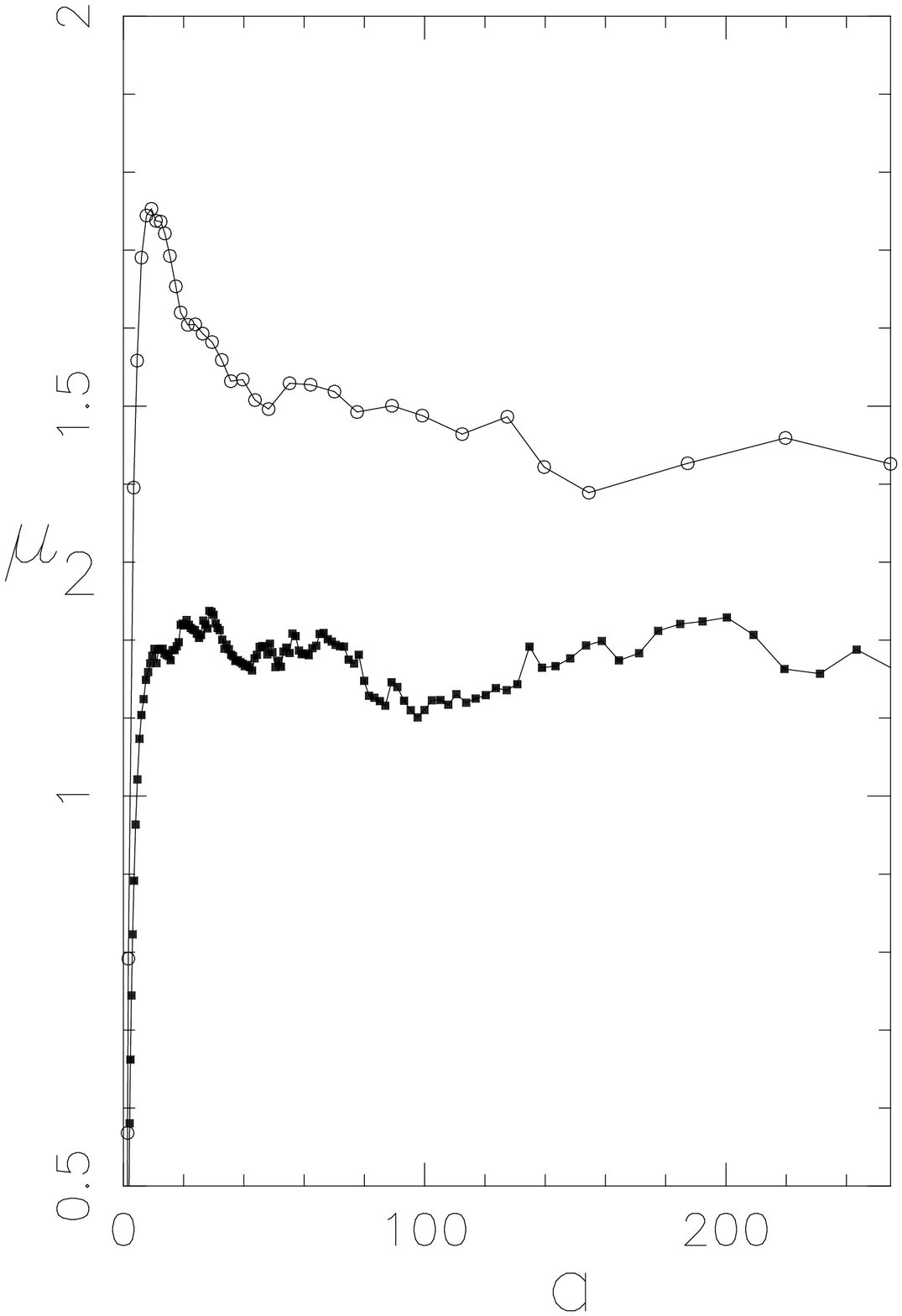,width=3.0 truein}}
\caption{ The second moment of the topological
distribution of the froth,$\mu_2$ vs. the mean area of the cells, for models
{\bf A} (full squares) and {\bf B} (circles). In all simulations the initial
configuration contains
1000 defects, placed randomly in a regular array of 40000 hexagons.
Each curve presents
data averaged over 6 runs.
The curve obtained using model {\bf B} exhibits a peak and
gives an asymptotic value in agreement with
the experimental one measured in the scaling state (1.4).
}
\label{fig:1}  \end{figure}

 \begin{figure}
\centerline{\psfig{file=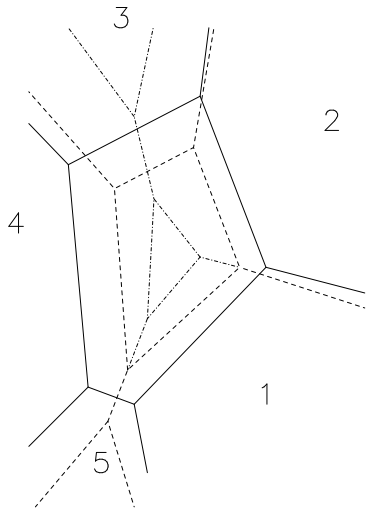,width=2.2 truein}}
\caption{
Disappearance of a pentagon
occurs in three stages:
{\it  Solid line:} just before
the T2-process; {\it  Dashed line:} after a T1 switch of the shortest
side the pentagon turns into a rectangle;
{\it  Dotted line:} after a T1 switch of its shortest side the square
becomes a triangle, which then
shrinks to a point.
}
\label{fig:2}
\end{figure}


\begin{references}
\bibitem{stor}
Note that only the indices of the $6N$ non-vanishing elements need to be
stored.
\end{references}
\end{document}